\documentclass[11pt,A4]{yic_2021_Extended_Abstract}

\usepackage{graphicx}
\usepackage{amsmath}
\usepackage{amsfonts}
\usepackage{amssymb}
\title{Computing the jump-term in space-time FEM for arbitrary temporal interpolation}

\author{Eugen Salzmann$^{*}$, Florian Zwicke$^{\dag}$ and Stefanie Elgeti$^{\dag}$}

\address{$^{*}$ Chair for Computational Analysis of Technical Systems(CATS)\\
RWTH Aachen University\\
Schinkelstr. 2, 52062 Aachen, Germany\\
e-mail: Salzmann@cats.rwth-aachen.de
\and
$^{\dag}$ Institute of Lightweight Design and Structural Biomechanics (ILSB)\\
Vienna University of Technology\\
Grumpendorferstr. 7, 1060 Vienna, Austria\\
e-mail:\{Zwicke,Elgeti\}@ilsb.tuwien.ac.at}

\abstract{One approach with rising popularity in analyzing time-dependent problems in science
and engineering is the so-called space-time finite element method that utilizes finite elements
in both space and time. A common ansatz in this context is to divide the mesh
in temporal direction into so-called space-time slabs, which are subsequently weakly
connected in time with a discontinuous galerkin approach. The corresponding jump-term,
which is responsible for imposing the weak continuity across space-time slabs, can
be challenging to compute, in particular in the context of deforming domains. Ensuring
a conforming discretization of the space-time slab at the top and bottom in time
direction simplifies the handling of this term immensely. Otherwise, a computationally
expensive and error prone projection of the solution from one time-level to another is
necessary. However, when it comes to simulations with deformable domains, e.g. for
free-surface flows, ensuring conforming meshes is quite laborious. A possible solution
to this challenge is to extrude a spatial mesh in time at each time-step resulting in the
so-called time-discontinuous prismatic space-time (D-PST) method \cite{Tezduyar90}. However, this
procedure is restricted to finite-elements of 1st order in time.
We present a novel algorithmic approach for arbitrarily discretized meshes by flipping
the mesh in time-direction for each time-step. This ansatz allows for a simple evaluation
of the jump-term as the mesh is always conforming. The cost of flipping the mesh
around its symmetry plane in time scales with the number of nodes, which makes it
computationally cheaper than an additional update of the mesh to enforce conformity or
the evaluation of a projection. We validate the approach on various physical problems
with and without deforming domains.}

\begin{document}
    
\section{INTRODUCTION}
Space-time is the extension of the finite element concept in time. It was first introduced in 1988 by Thomas J.R. Hughes for classical elastodynamics with a proven convergence theorem \cite{Hughes88}. Nowadays, it is more commonly used in fluid problems, especially since the introduction of the deformable-spatial-domain/space-time (DSD/SST) method \cite{Tezduyar90,Tezduyar90-2}. DSD/SST is beneficial for free-surface-flows, where the computational domain is unknown as they allow for a convenient way to track the boundary \cite{elget15}. The initial version of space-time, as well as many other adaptations, involves a discontinuous galerkin approach in time that leads to an additional jump term between so-called space-time slabs. The evaluation of this term can be challenging. Therefore we present an algorithmic approach for a more straightforward implementation of this term. 

\section{METHOD}
In this section, we present our approach to the jump term in DG-Space-time methods. The section is structured as follows. First, we recap the basics of space-time methods, including the treatment of deforming domains. The next part focuses on possible treatments of the jump-term, including the flipping ansatz.  

\subsection{SPACE-TIME}
\begin{figure}[ht]
    \label{fig:st_slab}
    \def\svgwidth{0.5\textwidth}
    \centering
    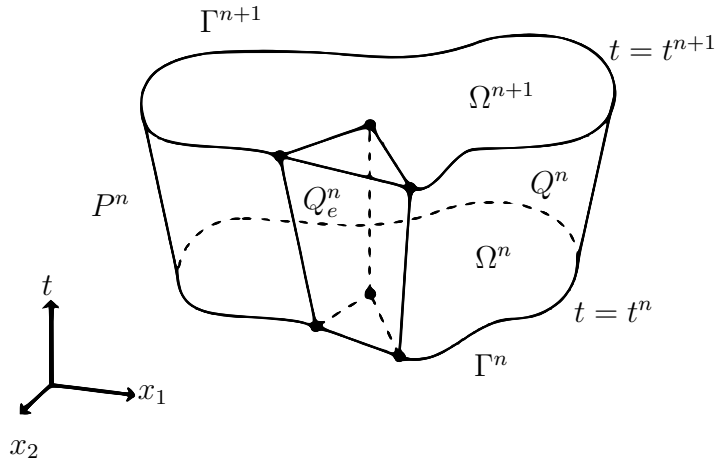
    \caption{Illustration of a space-time-slab $Q^n$ and an exemplary single element $Q_e$.}
\end{figure}

Space-time methods utilize finite elements in space and time rather than finite differences as in semi-discrete settings, resulting in a finite element analysis of the full space-time domain. There are various space-time methods that can be categorized with respect to the employed element type as well as the time continuity. In this work, we choose prismatic elements and a discontinuous galerkin approach in time, resulting in an analysis of so-called space-time slabs illustrated in Fig. \ref {fig:st_slab}.
These slabs can consist of one or multiple elements in time and can be considered an extension of a spatial mesh in time-direction. A jump-term, as typical for DG methods, weakly enforces continuity in time direction over multiple slabs. In this context, the weak form of a transient heat conduction equation reads: find $T \in S_a\left(Q^n\right)$ such that $\forall w \in S_t\left( Q^n\right)$:
\begin{equation}
    \int_{Q^n}w \frac{\partial T}{\partial t} \, \mathrm{d\bold{x}} = \int_{Q^n} w \alpha \triangle T \, \mathrm{d\bold{x}} + \int_{\Omega_n}w \left( T|^+_{t_n}-T|^-_{t_n} \right) \mathrm{d\bold{x}}.
    \label{eq:st_w_heat_conduction}
\end{equation}
Where $Q^n=\Omega \times \left[ t_n, t_{n+1}\right ] ,\; T|^{\pm}_{t_n} = \lim \limits_{\epsilon \to 0}T(t_n \pm \epsilon), \;\alpha \, \hat{=}$ thermal diffusivity and $S_a,S_t$ are the appropriate ansatz and testing spaces on $Q^n$. 
Problems involving deforming domains can profit from space-time methods despite introducing additional complexity or restrictions. One benefit is incorporating deformation by formulating the weak form over the deforming or deformed domain. 
That way, the movement or deformation is considered in the solution procedure without modifying governing equations. When the movement is known, this can be particularly useful as the solution could be found over the entire time interval in one step. In problems where the deformation of a spatial domain is unknown beforehand, space-time methods with DG in time face a challenge concerning the jump term. The solution at the top of the slab from the previous time-step has to be projected to the bottom of the slab in the current time-step, which is easy when the mesh is conforming. In that case, there is a direct relation between the degrees of freedom.
For non-conforming meshes, it is more challenging, and one possible solution namely a projection relying introduces an additional error. Furthermore, the domain itself may deform, resulting in entirely non-matching domains. This can be avoided by (1) mesh update schemes that ensure mesh conformity or (2) a restriction to single-element layers in time. 

\subsection{MESH INVERSION/FLIPPING}
We propose an alternative algorithmic approach, where the evaluation of the jump-term is easy to implement and works for arbitrary discretizations. Let us first focus on the implementation of the space-time method itself. Consider a 3D semi-discrete domain as compared to a 2D+time domain. Even though geometrically identical, algorithmically, one observes differences. These lie in the underlying operators in the PDE as well as the additional jump term. In terms of differential operators, in space-time approaches, the temporal derivative needs to be evaluated in a finite element sense whereas spatial derivatives need to be restricted to the spatial dimension only and can no longer be evaluated on the full domain.  We adapt the FE mapping between the reference and physical space to consider the changes in the spatial differential operator and to scale the input mesh with the time-step size. As a result, within the computational mesh, the time coordinates are usually contained in $[0,1]$ and the physical time is considered only through the mapping. Furthermore, for every space-time slab, the initial solution, from where the iterative solution scheme commences, is set to zero. Note that this entails that one can manipulate mesh coordinates without disturbing the solution process. We make use of this and invert the time coordinates. For every new time-step, before the iterative solution process starts, for every mesh node, we set the new t-coordinate $t^*$ as
\begin{equation}
 t^*=t_{max}-t + t_{min}.
 \end{equation}
Here, $t_{max/min}$ is the maximum/minimum value for $t$ in the slab, which are commonly $1$ and $0$. As a result, the vertices and corresponding degrees of freedom move from the bottom to the top and vice versa without moving in space, ensuring conformity at the slab interface.

\section{NUMERICAL STUDIES}
This section aims to validate the presented approach. We focus on transient heat conduction problems and complement our test cases with a hyperplastic solid mechanical bending problem. For showing that the approach is valid for moving domains, we present a test case with a prescribed motion.
\subsection{TRANSIENT HEAT ANALYSIS}
\begin{figure}[h]
 \centering
        \def\svgwidth{0.8\textwidth}
     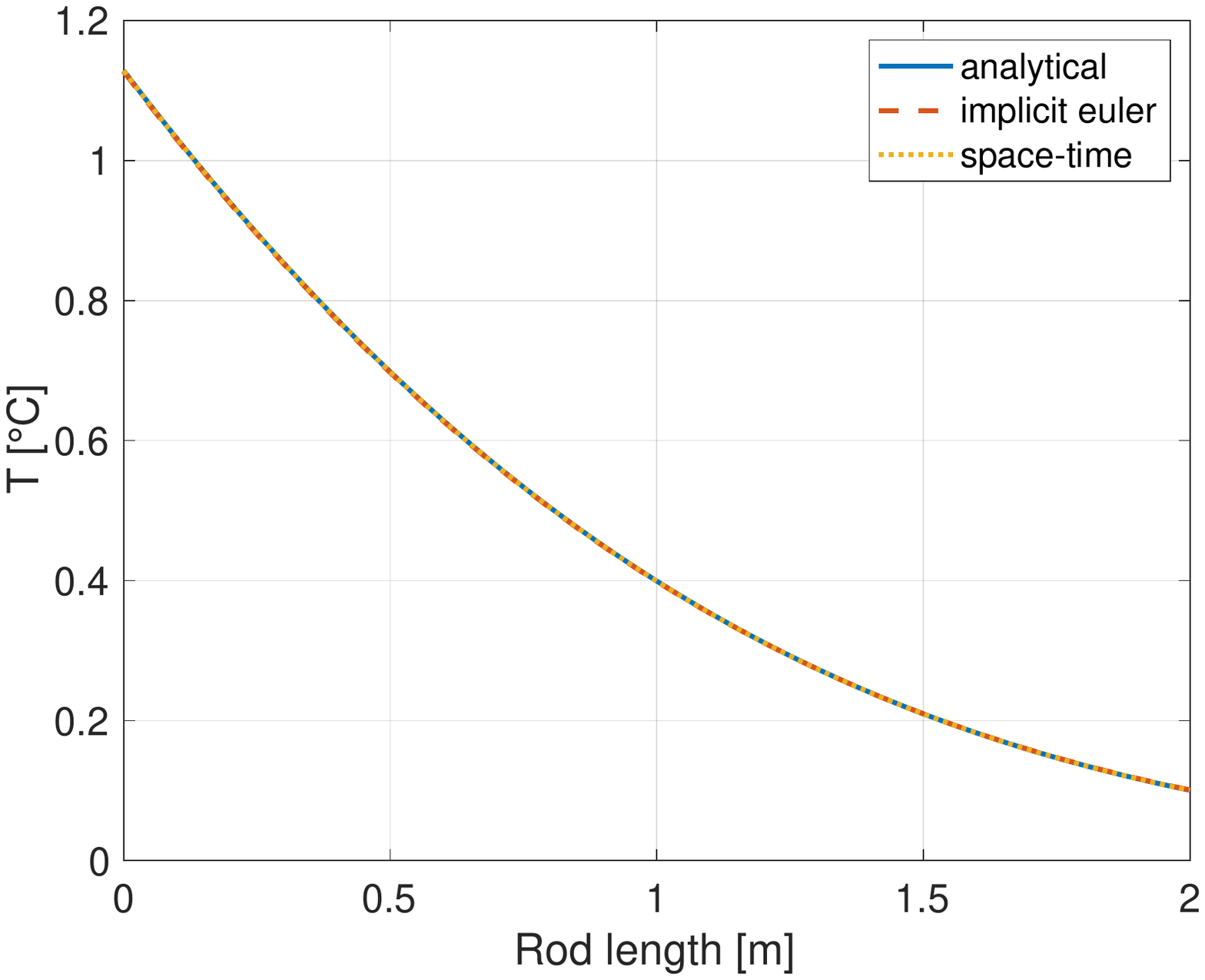
 \caption{Exact temperature distribution on the most left 2 m of the rod after 1 sec.}
  \label{fig:analytical_sol}
\end{figure}

\begin{figure}[h]
    \centering
    \def\svgwidth{0.8\textwidth}
    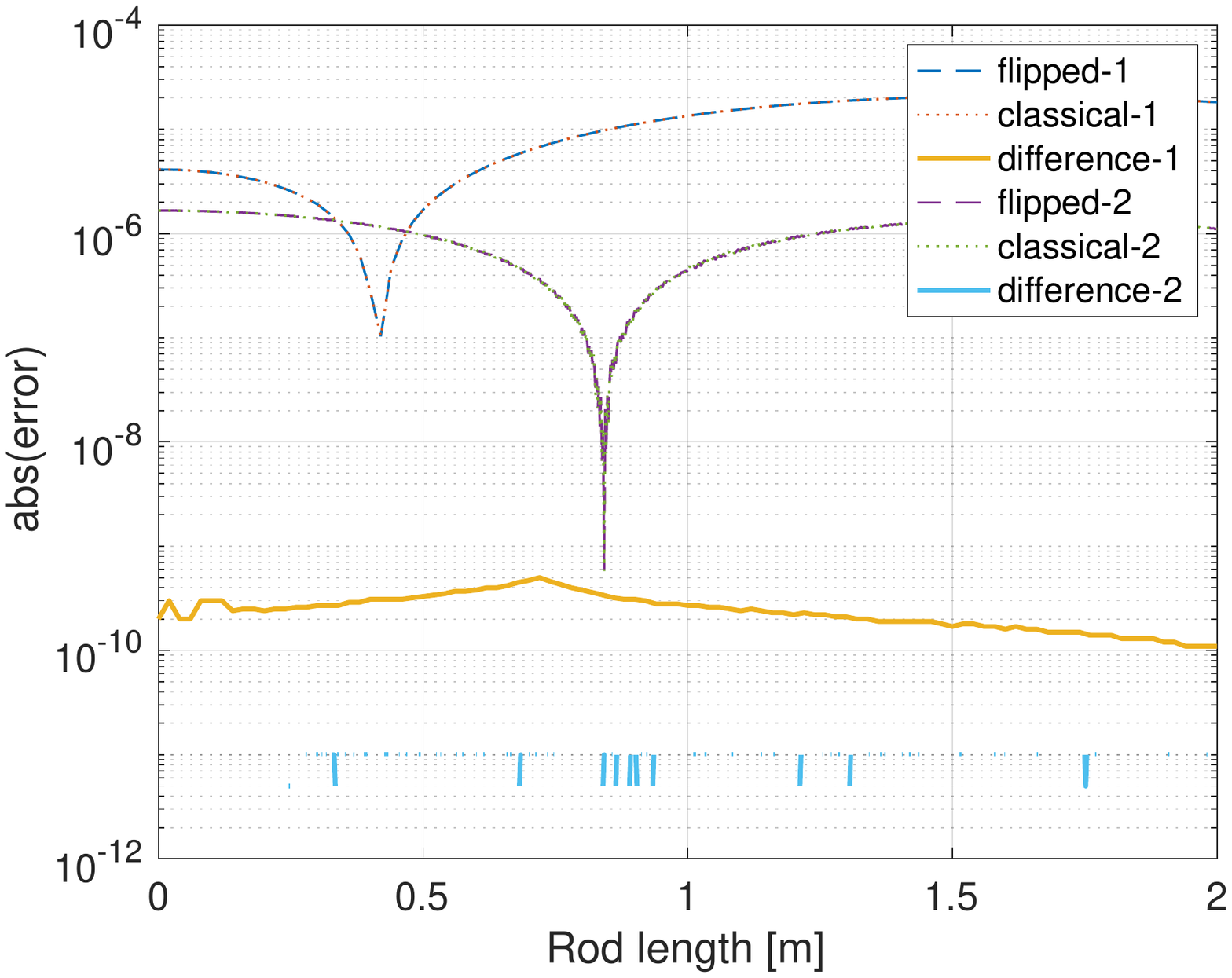
    \caption{Error comparison of heat conduction analysis between classical and flipped space-time.}
    \label{fig:dof_comparison}
\end{figure}

In our first test case, we follow \cite{Heat} and analyze heat conduction in a 2D rod. The rod is a rectangular domain and adiabatic everywhere except on the left side, where a fixed heat flux of 1 W is prescribed. 
\begin{table}[h]
    \centering
    \caption{Parameters for transient heat conduction analysis of a 2D rod.}
    \label{tab:rod_param}
    \begin{tabular}{ c| c| c}
       Parameter & Value & Unit \\
        \hline
        \hline
        Length & 20 &m \\
        \hline
        Width & 1 &m \\
        \hline
        Thermal diffusivity ($\alpha = \frac{\kappa}{\rho c_p}$)& 1 & $\frac{m^2}{s}$ \\
        \hline
        Reference time & 1 & s \\
        \hline
        Reference length & 2 & m
    \end{tabular}
\end{table}
Tab. \ref {tab:rod_param} contains the geometry and simulation parameters for this test case. Reference time and length refer to the values at which we compare with the exact solution. Despite the 2D geometry, the problem reduces to a 1D phenomenon as the heat propagates equally over the width. An analytical solution is given by
\begin{equation}
 \label{eq:rod_exact}
 T\left(x,t\right) = 2\sqrt\frac{t}{\pi}\left[ e^{\frac{-x^2}{4t}} -\frac{1}{2}x\sqrt\frac{\pi}{t} erfc\left(\frac{x}{2\sqrt t}\right)\right],
\end{equation}
and shown in Fig. \ref{fig:analytical_sol} together with a space-time and implicit Euler solution. The resulting solutions for the temperature distribution are depicted in Fig. \ref{fig:analytical_sol}. All solutions are visually identical. Therefore, we proceed to analyze the errors of the individual implementations.
Our first comparison is between the presented approach and a classical implementation of the space-time jump-term. Fig. \ref{fig:dof_comparison} shows the error between these techniques and between each of them and the exact solution for two different discretizations. 
The curve index refers to the discretization while the "difference" curves show the error between the two implementations. The mesh details are the following: 
\begin{table}[h!]
    \centering
    \caption{Mesh sizes for comparison between flipped and classical space-time implementations.}
    \label{tab:mesh_dof_comp}
    \begin{tabular}{ c| c| c | c}
        Curve  index & elements in length & elements in width & time step size \\
        \hline
        \hline
        1& 1000 & 1 & 0.1 \\
        \hline
        2 & 10000 & 1 & 0.05\\
    \end{tabular}
\end{table}
\\
 From Fig. \ref{fig:dof_comparison}, it is evident that the effect of the mesh inversion is negligable in comparison to the overall discretization error.
 \begin{figure}[h!]
     \centering
     \def\svgwidth{0.8\textwidth}
     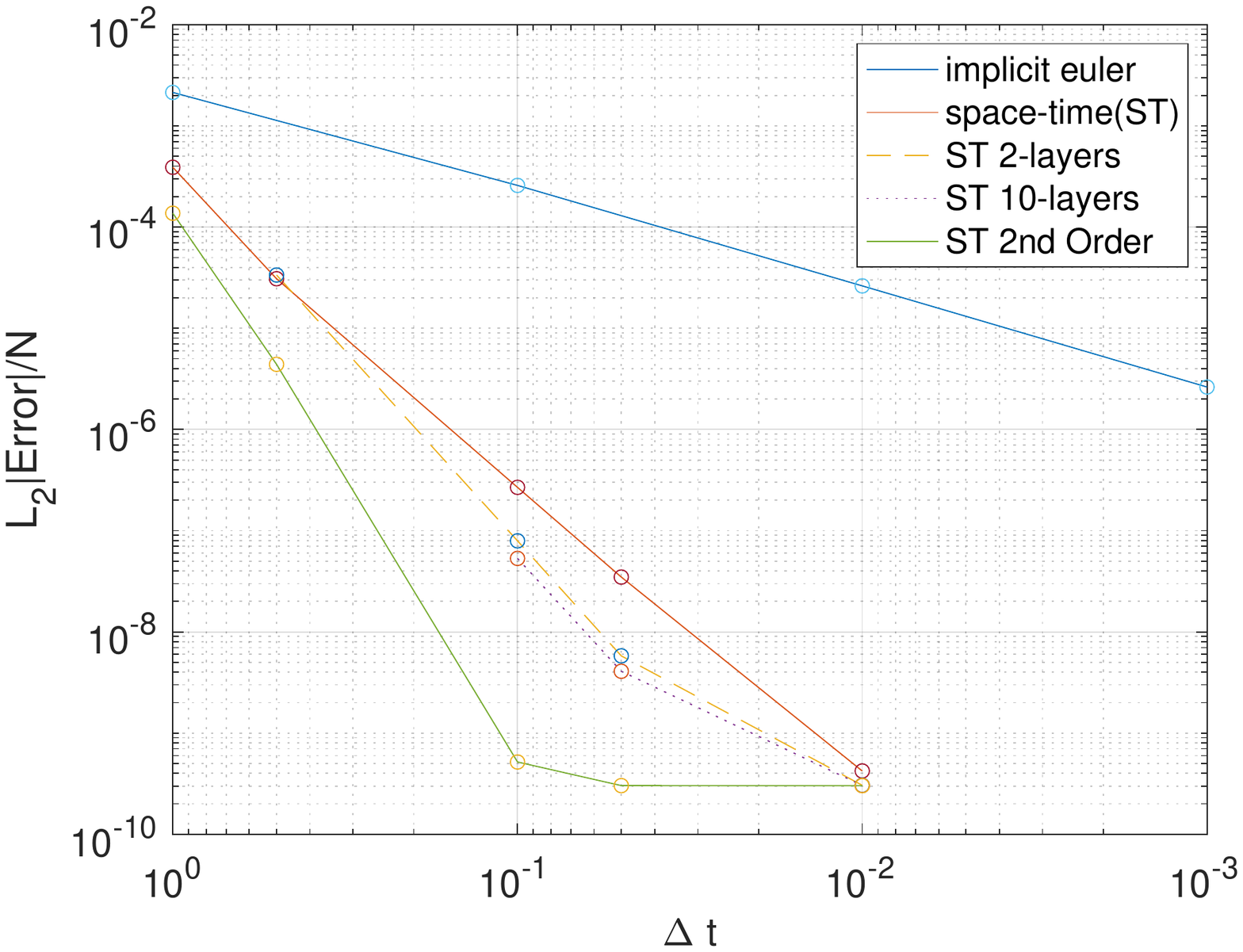
     \caption{Relative error evolution of heat conduction analysis under temporal refinement.}
     \label{fig:convergence}
 \end{figure}
 Fig. \ref{fig:convergence} shows the evolution of the error between the analytical solution and various space-time simulations as well as one implicit Euler simulation under temporal refinement. The error is evaluated as the L-2 norm over the first 2 m of the mesh normalized by the number of evaluation points (N). The analysis was performed on a quadrilateral 100.000x1 element spatial mesh. Note that the error of the implicit Euler scheme comes close to the theoretical convergence order of one. In comparison, the error of the space-time ansatz decreases significantly faster when decreasing the time step size. However, not further than about $5.0e^{-10}$ where the spatial discretization error dominates and temporal refinement results in no improvement. For a fair comparison, the depicted time-step size $\Delta t$ in the case of multiple time layer meshes corresponds to the layer thickness. 
 
 
\subsection{STRUCTURAL ANALYSIS}
 \begin{figure}[h]
    \centering
    \def\svgwidth{0.6\textwidth}
    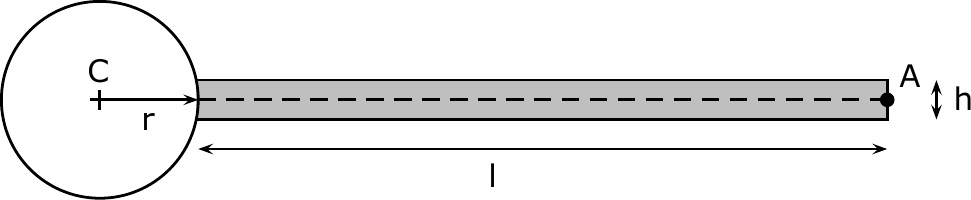
    \caption{Illustration of the beam used for structural analysis \cite{Turek1}.}
    \label{fig:beam_geo}
\end{figure}
\begin{figure}[h!]
    \centering
    \def\svgwidth{0.8\textwidth}
    \input{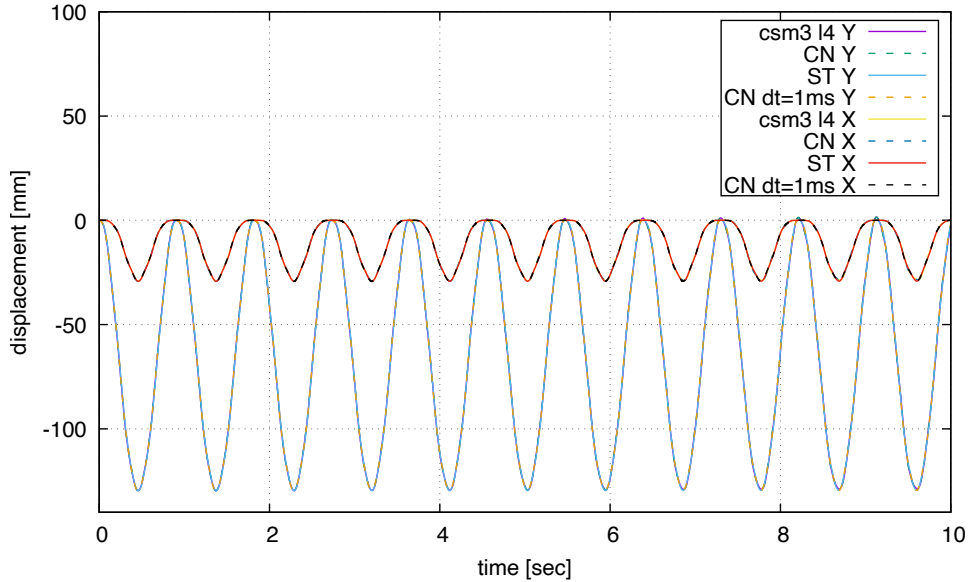}
    \caption{Displacement of reference point A over 10 seconds.}
    \label{fig:beam_full}
\end{figure}
\begin{figure}[h!]
    \centering
    \def\svgwidth{1\textwidth}
    \input{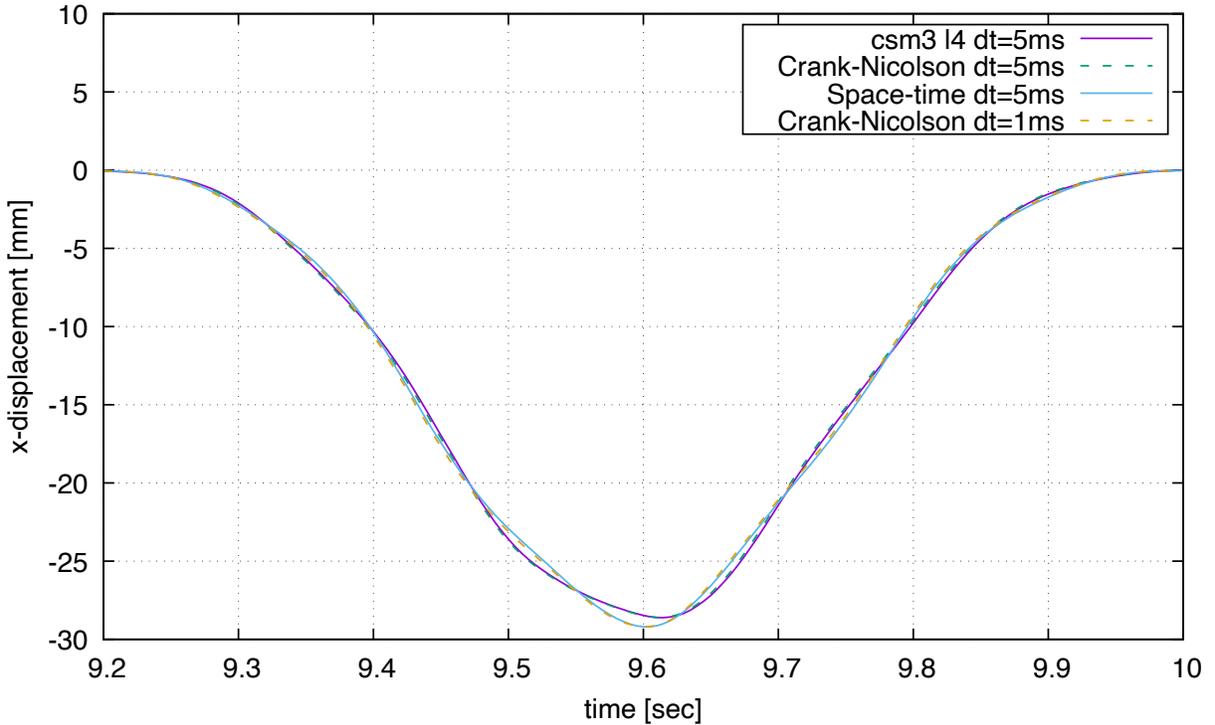}
    \caption{Displacement in x-direction within the last considered period.}
    \label{fig:beam_9_x}
\end{figure}

The next test case we would like to present is the transient structural analysis of a beam that is fixed on the left side to a cylinder and bending under its weight. The clylinder is not part of the computational domain. The test case is taken from \cite{Turek1} where it is used to validate a structural solver before investigating fluid-structure-interaction. A hyperelastic material model, namely the St-Venant-Kirchhoff model, is employed. The problem is modeled from a classical lagrangian point of view, and we avoid 2nd order temporal derivatives by introducing a velocity field and solving it. The geometry is depicted in Fig. \ref{fig:beam_geo}, and the following parameters were used:
\vspace{-0.5cm}
\begin{table}[h!]
    \centering
    \caption{Parameters of structural analysis test-case.}
    \label{tab:parm_beam}
    \begin{tabular}{ c| c| c }
       Parameter & value & unit  \\
        \hline
        \hline
        length& 35 & cm \\
        \hline
        width & 2 & cm \\
        \hline 
        cylinder radius $r$ & 5 & cm   \\
        \hline
        density & 1000 & $\frac{kg}{m^3}$ \\
        \hline 
        1st lamme parameter $\lambda$  & 2 & $10^6\frac{kg}{ms^2}$  \\
        \hline 
        2nd lamme prameter $\mu$ & 0.5 &$10^6\frac{kg}{ms^2}$  \\
        \hline
        gravity (y-direction)& -2 & $\frac{m}{s^2}$ \\ 
    \end{tabular}
\end{table}
\\
We compare the displacement of the reference point A at the tip of the beam shown in Fig. \ref{fig:beam_geo}. Fig. \ref{fig:beam_full} illustrates the displacement over 10 seconds in X and Y direction. The simulation details are given in Tab. \ref{tab:mesh_parm_beam}. The results seem to be in good agreement. 
\begin{table}[h!]
    \centering
    \caption{Simulation parameters of structural analysis results.}
    \label{tab:mesh_parm_beam}
    \begin{tabular}{ c| c| c }
        run & elements & $\Delta t$ \\
        \hline
        \hline
        csm l4 & 5120 & 5ms  \\
        \hline
        Crank-Nicolson (CN) & 20x128 & 5ms \\
        \hline 
        Space-time (ST)&20x128  & 5ms   \\
        \hline
        CN dt=1ms& 20x128 & 1ms \\
    \end{tabular}
\end{table}
Nevertheless, there are minor deviations. Fig. \ref{fig:beam_9_x} zooms in on the last period of the oscillation between 9.2 and 10 seconds, and we focus on the displacement in x-direction as the differences are more visible in that direction. The result employing a Crank-Nicolson scheme and the same time-step is very close to the reference values, and different meshes can explain the occurring fine distinctions. ‘CSM l4’ uses an unstructured mesh with 5120 quadrilateral elements, while we choose a structured mesh of 20x128 quadrilateral elements. The space-time result exhibits slightly different behavior, which is close to the results of a Crank-Nicolson simulation with a refined time-step. However, as shown in the previous test case, space-time exhibits a superior accuracy with respect to the time-step size, so this result is to be expected.
\subsection{RIGID-BODY MOVEMENT}
\begin{figure}[h]
    \centering
    \def\svgwidth{1\textwidth}
    \input{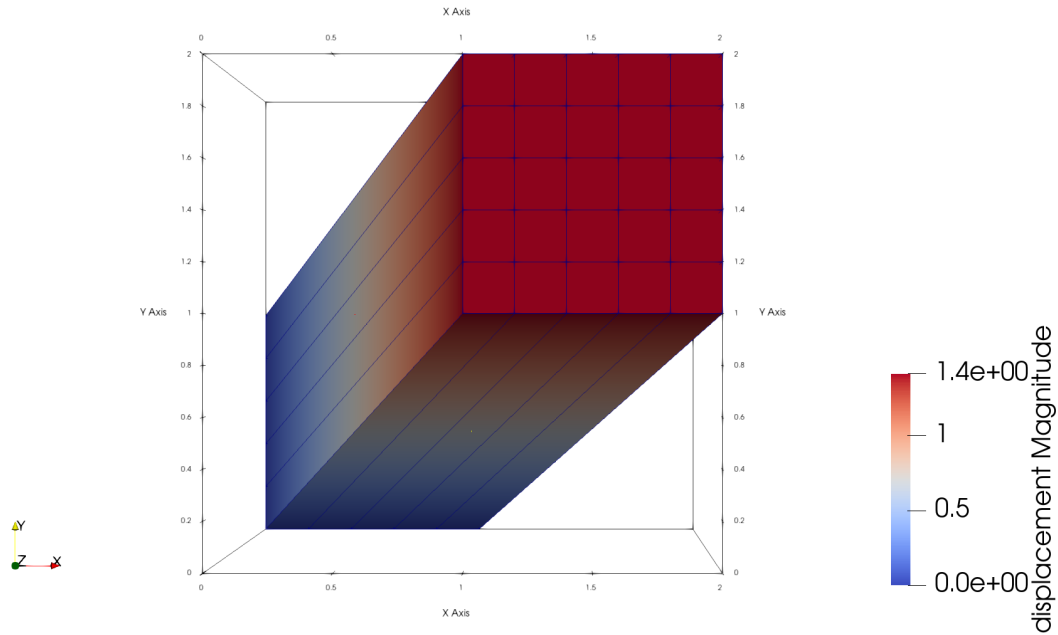}
    \caption{Rigid-body movement result after 10 seconds.}
    \label{fig:move}
\end{figure}

The last test case we are discussing is the movement of a rigid body. This test case is somewhat artificial as we solve a continuum mechanical system of equation coupled with an elastic mesh-update problem and a free-surface approach, similar to what is described by Elgeti and Zwicke \cite{elget15,Zwicke17}. However, as we are only interested in the deformation of the domain, we reduce the problem to a rigid body movement by imposing a fully developed velocity field of $ v=(0.1,0.1)^T$ as initial and boundary conditions. The test case is a 1x1 sized spatial domain that then has to move $(1,1)^T$ in 10 seconds. Fig. \ref{fig:move} shows the result after one time-step with size 10. The t-axis is pointing out of the plane, and we view the x-y plane, meaning, at the bottom, the body is in its initial configuration, and the mesh connects it to its new position after 10 seconds at the top. The legend shows that the magnitude of the displacement is $\sqrt{2}$ after 10 seconds, which corresponds to the analytical solution.

\section{CONCLUSION}
In this work, we presented an algorithmic approach to the treatment of jump terms in the context of space-time discontinuous finite element methods. We showed that inverting the space-time slab around its temporal axis leads to a one-to-one degree of freedom correspondence on the new bottom of the slab, allowing for easy evaluation of the jump term. Additionally, this alleviates the requirement of a conforming mesh on the top and bottom of the time-slab introducing new flexibility for the discretization. This ansatz is especially advantageous for problems involving moving domains, where the movement is not known apriori. Our numerical studies validated the approach in comparison to classical space-time and semi-discrete FEM solutions and analytical solutions for different physical problems.

\end{document}